\documentclass[letterpaper,
twocolumn,
aps,nofootinbib,superscriptaddress,tightenlines,10pt,notitlepage,
pra,
reprint
]{revtex4-1}
\pdfoutput=1

\usepackage{amscd,amsfonts,amsmath,amssymb,amsthm,dsfont,mathrsfs,mathtools}
\usepackage{color,graphicx,enumerate}
\usepackage{natbib}
\usepackage[
colorlinks=true, 
linkcolor=black, 
citecolor=blue, 
urlcolor=blue, 
hyperindex,
]{hyperref}

\usepackage{color}

\usepackage[utf8]{inputenc}
\usepackage[T1]{fontenc}
\usepackage{lmodern}
\usepackage{centernot}

\usepackage{blindtext}

\usepackage{tikz}
\usetikzlibrary{arrows}

\theoremstyle{plain}
\newtheorem{theorem}{Theorem}

\newtheorem{property}[theorem]{Property}

\theoremstyle{definition}

\def\:{\colon}

\DeclareMathOperator{\Tr}{Tr}

\providecommand{\abs}[1]{\lvert#1\rvert}

\providecommand{\sizedabs}[1]{\left\lvert#1\right\rvert}

\providecommand{\ket}[1]{\lvert#1\rangle}

\providecommand{\braket}[2]{\langle#1|#2\rangle}
\providecommand{\ketbra}[2]{\lvert#1\rangle\!\langle#2\rvert}

\def\Gammaminus{{\Gamma_{\!-}}}
\def\Gammaplus{{\Gamma_{\!+}}}

\begin{document}

\title{The binegativity of two qubits}
\date{\today}
\author{Mark W.\ Girard}
\email{mwgirard (at) ucalgary.ca}
\affiliation{Department of Mathematics and Statistics and Institute for Quantum Science and Technology, University of Calgary\\ 2500 University Dr NW, Calgary, AB T2N 1N4, Canada}
\author{Gilad Gour}
\affiliation{Department of Mathematics and Statistics and Institute for Quantum Science and Technology, University of Calgary\\ 2500 University Dr NW, Calgary, AB T2N 1N4, Canada}

\begin{abstract}
We introduce an entanglement-related quantity that we call the binegativity. Based on numerical evidence, we conjecture that the binegativity is an entanglement measure for two-qubit states. The binegativity is compared to the concurrence and negativity, the only known entanglement measures that can be computed analytically for all two-qubit states. The ordering of entangled two-qubit mixed states induced by the binegativity is distinct from the ordering determined by previously known measures, yielding possible new insights into the structure of entangled two-qubit mixed states. 
\end{abstract}

\maketitle


Entanglement is a distinguishing feature of quantum mechanics and is a key ingredient in many quantum information processing tasks. Substantial progress has been made towards classifying quantum entanglement in the past two decades \cite{Bennett1996b,Horodecki2009}, but much remains unknown even in the case of bipartite states. Entanglement of pure states is well understood \cite{Bennett1996c,Nielsen1999,Vidal2000,Lo2001,Vedral2002,Vidal2003}, but despite the enormous amount of work done on  mixed-state entanglement, many fundamental challenges still remain \cite{Horodecki2009}.

Two primary goals of entanglement theory include: determining whether a given state is entangled, and finding conditions for when two states can be converted under local operations and classical communication (LOCC) \cite{Plenio2014,Girard2015a,Plenio2007}. Conditions for convertibility of entangled states are typically determined by quantifying entanglement in terms of entanglement measures---quantities that do not increase under application of  LOCC operations \cite{Plenio2007,Vidal2000a,Eisert1998,Vedral1997}. Finding new entanglement measures is important for furthering our understanding of the structure of entangled states. While there are numerous known quantifications of bipartite entanglement \cite{Horodecki2009}, there are only two entanglement measures that are known to be effectively computable for all states of two qubits---the concurrence \cite{Hill1997,Wootters1998} and the negativity \cite{Vidal2002}. The negativity coincides with the concurrence for two-qubit pure states, but these entanglement measures reveal different orderings on two-qubit mixed states \cite{Miranowicz2004}. 

One important mathematical tool in entanglement theory is the partial transpose operation \cite{Peres1996}, which can be used to detect entanglement in states. The partial transpose of a state $\sigma$ is denoted $\sigma^\Gamma$. States satisfying $\sigma^\Gamma\geq 0$ are said to be positive under partial transposition (PPT). It is known that all separable states are PPT \cite{Peres1996} and all PPT states of two-qubits are separable \cite{Horodecki1996}. The negativity quantifies the degree to which a state violates the PPT criterion. Although there is no known physical interpretation for the partial transpose, the negativity is an entanglement monotone \cite{Vidal2002}. The negativity can be written as
\begin{equation*}
 N(\sigma) = 2\Tr[\sigma^{\Gammaminus}],
 \end{equation*}
where we use the notation 
\begin{equation*}
 \sigma^{\Gammaplus}:=(\sigma^\Gamma)_+ \quad\text{ and }\quad \sigma^{\Gammaminus}:=(\sigma^\Gamma)_-
\end{equation*}
to denote the positive and negative components of the partially transposed $\sigma$, respectively. The negativity is non-increasing under both deterministic LOCC channels \cite{Plenio2005} and the larger class of PPT channels \cite{Rains2001}. 
 
In this paper we introduce a computable quantity that we call \emph{binegativity} given by
 \begin{equation}
 N_2(\sigma):=\Tr[\sigma^{\Gammaminus}] + 2\Tr[\sigma^{\Gammaminus\Gammaminus}],
 \end{equation}
where $\sigma^{\Gammaminus\Gammaminus} = ((\sigma^\Gammaminus)^\Gamma)_-$. This is based on a construction given in Ref.\ \cite{Girard2015a}.  We show that the binegativity has similar properties to the negativity. While we do not have a valid proof that it behaves monotonically under both LOCC and PPT channels for states of two-qubits, we suspect that it is based on numerical evidence. That is, we conjecture that for any two-qubit state $\sigma$ it holds that
\[
 N_2(\mathcal{E}(\sigma))\leq N_2(\sigma)
\]
for any LOCC (or PPT) channel $\mathcal{E}$ that outputs states of two qubits. The binegativity is also easily computable in that it can be deduced from the eigenvalue decomposition of the partial transpose of the state. It is also invariant under local unitaries and vanishes exactly on the separable states of two qubits. The remainder of this paper is dedicated to presenting key features of the binegativity. 

Since we conjecture that the binegativity is an entanglement measure, it is important to compare the binegativity to both the negativity and the concurrence, the only known computable measures of entanglement for arbitrary two-qubit states. The concurrence and the negativity give two different orderings for states of two-qubits \cite{Miranowicz2004}. The binegativity is closely related to the standard negativity, but we will show that it is distinctly different. We also show that there exist states that have equal negativity and equal concurrence but different values for the binegativity. Therefore, if the binegativity is in fact an entanglement measure, the binegativity gives a completely new ordering on the set of entangled two-qubit states that is different from the ordering determined by the negativity and concurrence. This would give us new insight into the structure of entangled two-qubit mixed states should monotonicity of the binegativity under PPT operations prove to hold.
%
\\

\noindent\textbf{\emph{Properties of and bounds for the binegativity}} ---
We first state and prove a few properties relating the binegativity to the negativity and concurrence. 

\begin{property}
For all two-qubit states, $N_2(\sigma)=0$ if and only if $\sigma$ is separable. 
\end{property}

\begin{proof}
 Let $\sigma$ be a state of two-qubits. If $\sigma$ is separable then $\sigma^\Gammaminus=0$ and thus $\sigma^{\Gammaminus\Gammaminus}=0$. On the other hand, if $N_2(\sigma)=0$ then $\sigma^\Gammaminus=0$ and thus $\sigma$ is separable. 
\end{proof}

\begin{property}
 For all two-qubit states $\sigma$, it holds that
 \begin{equation}
  N_2(\sigma)\leq N(\sigma)\leq C(\sigma)
 \end{equation}
 and $N_2(\sigma)=N(\sigma)$ if and only if $N(\sigma)=C(\sigma)$. 
\end{property}
\begin{proof}
 Let $\sigma$ be a state of two qubits. It is known \cite{Verstraete2001} that $N(\sigma)\leq C(\sigma)$ with equality if and only if the eigenvector of $\sigma^\Gamma$ corresponding to its negative eigenvalue is maximally entangled. Since the negativity and concurrence are both faithful entanglement measures for two-qubit states \cite{Verstraete2001a}, we may suppose that $\sigma$ is entangled. The partial transpose of any entangled two-qubit state has at exactly one negative eigenvalue \cite{Verstraete2001a,Verstraete2001}, and the pure state corresponding to that eigenvalue must be entangled. The negative component of $\sigma^\Gamma$ therefore has the form
\begin{equation}\label{eq:2qsigmagamminus}
 \sigma^{\Gammaminus} = \Tr[\sigma^{\Gammaminus}] \ketbra{\psi}{\psi},
\end{equation}
for some pure entangled state $\ket{\psi}$, where $\Tr[\sigma^{\Gammaminus}]=\frac{1}{2}N(\sigma)>0$. We may suppose without loss of generality that $\ket{\psi}$ is in Schmidt form
\[
 \ket{\psi} = \sqrt{\mu}\ket{00} + \sqrt{1-\mu}\ket{11}
\]
for some $\mu\in[\frac{1}{2},1)$. Then 
\begin{equation}
 \sigma^{\Gammaminus\Gammaminus}=\sqrt{\mu(1-\mu)}\Tr[\sigma^{\Gammaminus}]\ketbra{\psi^-}{\psi^-},
\end{equation}
where $\ket{\psi^-}=(\ket{00}-\ket{11})/\sqrt{2}$ is a maximally entangled Bell state. Hence $\Tr[\sigma^{\Gammaminus\Gammaminus}]\leq \frac{1}{2}\Tr[\sigma^{\Gammaminus}]=\frac{1}{4}N(\sigma)$ since $\sqrt{\mu(1-\mu)}\leq \frac{1}{2}$. From the definition of the negativity, it follows that $N_2(\sigma)\leq N(\sigma)$ with equality if and only if $\mu=\frac{1}{2}$ in which case $\ket{\psi}$ is maximally entangled. This proves the desired result.
\end{proof}

In particular, we see that $N_2(\psi)=N(\psi)=C(\psi)$ for all two-qubit pure states $\ket{\psi}$. Also note that $N_2(\sigma)=N(\sigma)=C(\sigma)=0$ for all two-qubit separable states \cite{Miranowicz2004}. We also point out that all three of these quantities vanish exactly on the separable state for two qubits.

The negativity and the concurrence have been compared for two-qubit states~\cite{Eisert1998,Verstraete2001,Verstraete2001a,Miranowicz2004}, and these entanglement measures give different orderings of states, since there exist states with equal negativity but different concurrence (and vice versa). The range of possible values for the negativity of two-qubit states with fixed concurrence $C(\sigma)=c$ is known \cite{Verstraete2001a} to be 
\begin{equation}\label{eq:negcbounds}
 \nu_c:= \sqrt{(1-c)^2 + c^2}-(1-c)\leq N(\sigma) \leq c. 
\end{equation}
The upper bound of \eqref{eq:negcbounds} is obtained by pure states while the lower bound is obtained by the family of two-qubit states of the form \cite{Verstraete2001a,Ishizaka2004}
 \begin{equation} \label{eq:sigMEMS}
  \sigma_c := c\ketbra{\phi^+}{\phi^+} + (1-c)\ketbra{10}{10}
 \end{equation}
for $c\in[0,1]$. These states have concurrence $C(\sigma_c)=c$ and negativity given by $N(\sigma_c)=\nu_c$, where $\nu_c$ is as defined in \eqref{eq:negcbounds}. The binegativity of these states can also be computed and simplifies to
\begin{equation}\label{eq:N2mems}
 N_2(\sigma_c) = \frac{\nu_c}{2}\left(1+\frac{c}{\sqrt{(1-c)^2+c^2}}\right).
\end{equation}

\begin{figure}
 \includegraphics[width=\columnwidth]{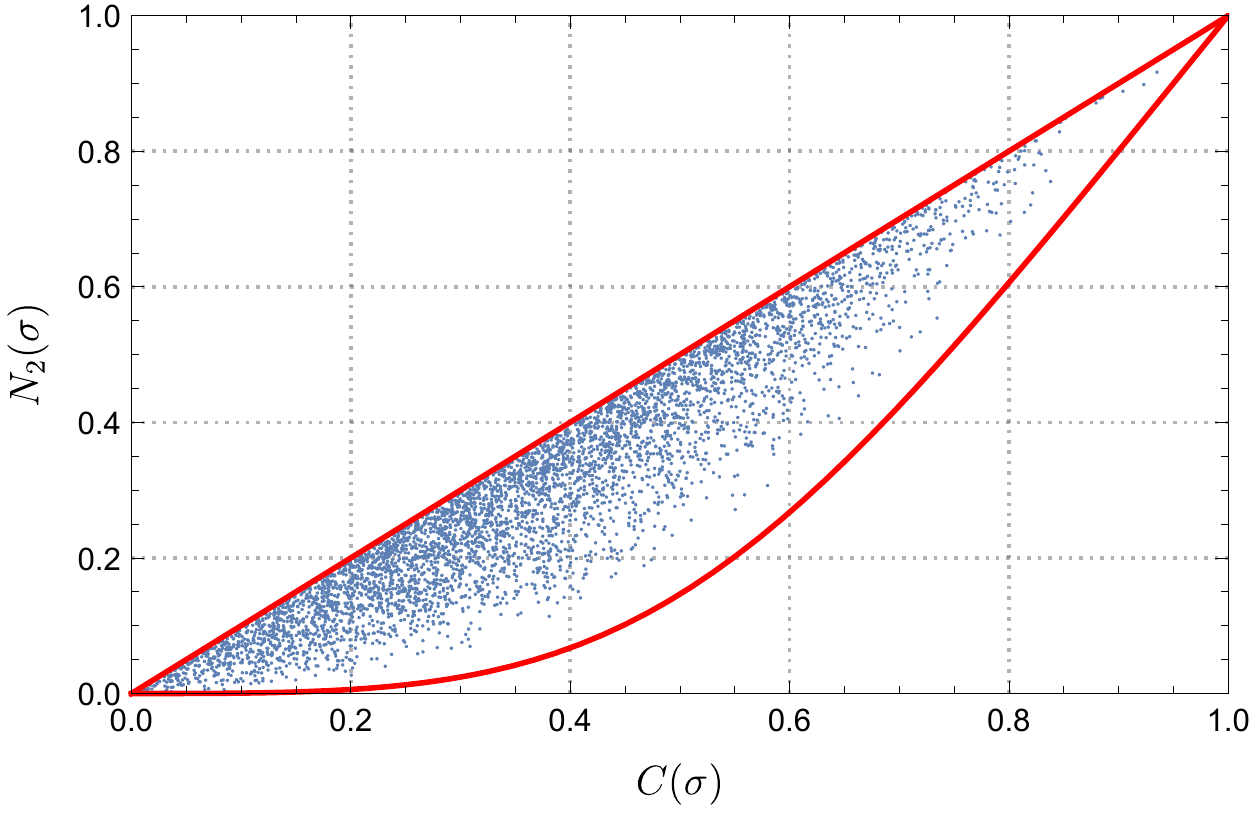}
 \caption{Concurrence vs binegativity for randomly generated two-qubit states of rank 2. The concurrence $C(\sigma)$ is on the horizontal axis and $N_2(\sigma)$ is on the vertical axis. The upper bound and (conjectured) lower bound are indicated by the solid red curves. The binegativity is bounded above by $C(\sigma)$ and the conjectured lower bound is given in \eqref{eq:N2mems}.}
 \label{fig:convsbinegqubit}
\end{figure}

We now compare the binegativity to the negativity and concurrence for randomly generated states of two qubits. A comparison of the concurrence and the binegativity for randomly generated states is shown in Fig.~\ref{fig:convsbinegqubit}. The binegativity is always bounded above by $N_2(\sigma)\leq C(\sigma)$, and this bound is achieved for pure states. Based on numerical evidence, we conjecture that the lower bound of the binegativity is given by \eqref{eq:N2mems}, where $c=C(\sigma)$. That is, the states that have minimal binegativity for a fixed value of the concurrence are the states in \eqref{eq:sigMEMS}.

\begin{figure}
 \includegraphics[width=\columnwidth]{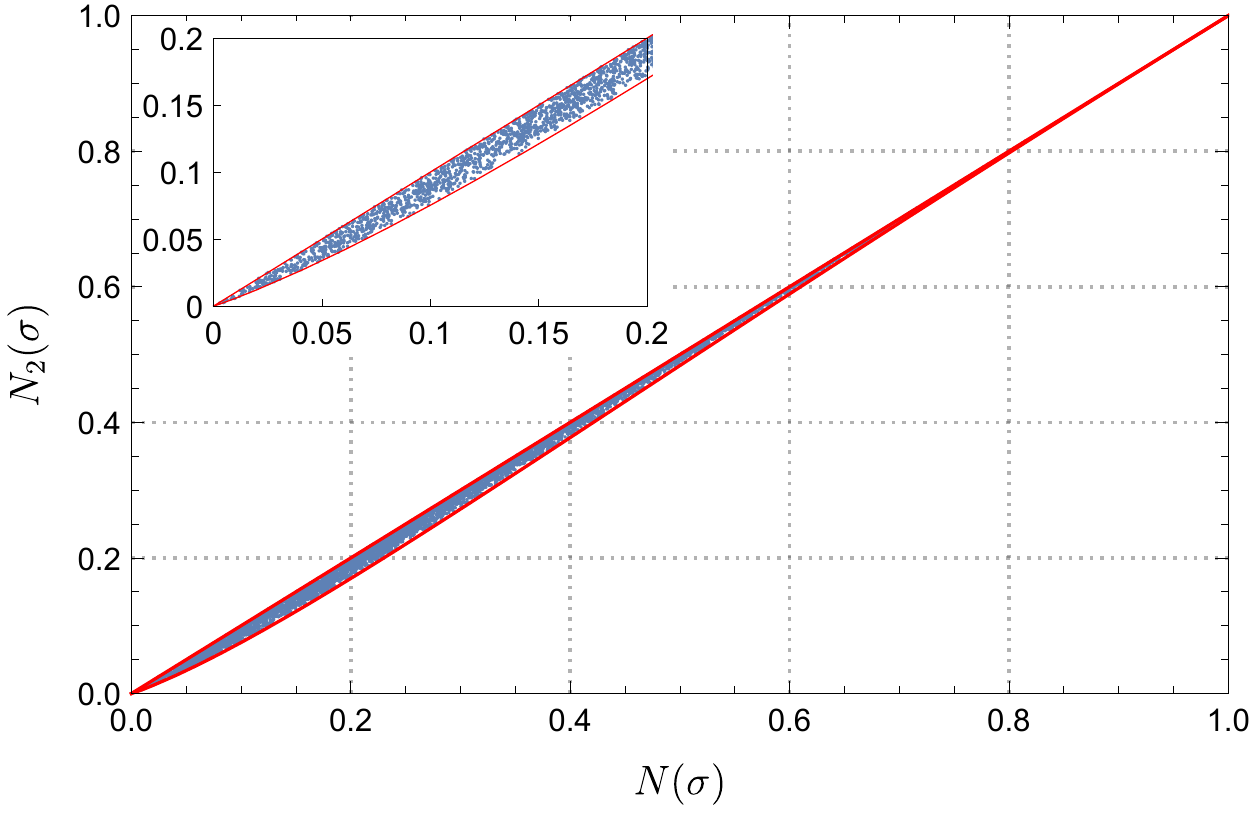}
 \caption{Negativity vs binegativity for randomly generated two-qubit states of rank 2. The negativity $N(\sigma)$ is on the horizontal axis and the binegativity $N_2(\sigma)$ is on the vertical axis. The upper bound and (conjectured) lower bound are indicated by the solid red curves. The binegativity is bounded above by $N(\sigma)$ and the conjectured to be bounded below by \eqref{eq:binegneglowerbound}, which is the binegativity of the states in \eqref{eq:sigMEMS}.}
 \label{fig:negvsbinegqubit}
\end{figure}

Next we compare the binegativity to the negativity of randomly generated two-qubit states, as shown in Fig.\ \ref{fig:negvsbinegqubit}. The binegativity is always bounded above by $N_2(\sigma)\leq N(\sigma)$ and this bound is achieved for pure states. Based on numerical evidence, we conjecture that the binegativity is bounded below by 
\begin{equation}\label{eq:binegneglowerbound}
  \frac{\nu}{2} \left(1+\frac{c_\nu}{\sqrt{(1-c_\nu)^2+c_\nu^2}}\right) \leq N_2(\sigma)
\end{equation}
where we define $c_\nu=\sqrt{2\nu(\nu+1)}-\nu$. Note that $c_\nu$ is the concurrence of the state of the form in \eqref{eq:sigMEMS} that has negativity given by $N(\sigma_{c_\nu})=\nu$. 

\begin{figure}
 \includegraphics[width=\columnwidth]{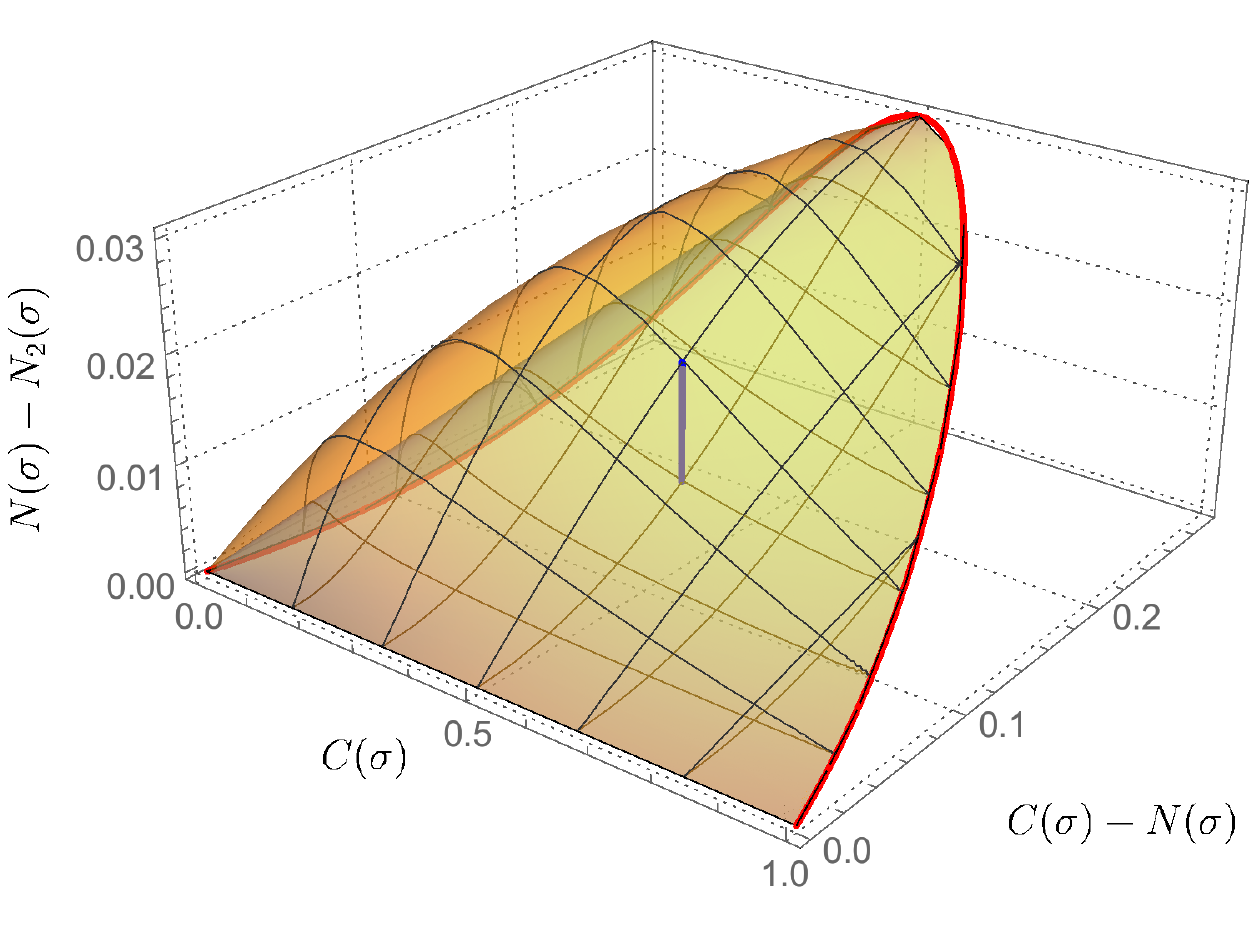}
 \caption{Conjectured region for the possible values of the concurrence, negativity, and binegativity for two-qubit states. The concurrence $C(\sigma)$ is on the $x$-axis, while the difference $C(\sigma)-N(\sigma)$ and $N(\sigma)-N_2(\sigma)$ are plotted on the $y$- and $z$-axes respectively. The upper and lower surfaces of the depicted region are given by \eqref{eq:bnegbounds}. While not shown in this figure, all randomly generated states of two-qubits were found to have binegativity within this region, and we conjecture that all two-qubit states have binegativity in the depicted region. The states of the form in \eqref{eq:sigMEMS} are depicted as the red curve. The vertical blue line in the centre connects the two states defined in \eqref{eq:rho12} that have equal concurrence and equal negativity but different binegativity.}
 \label{fig:cvsnvsb}
\end{figure}

It is also possible to simultaneously compare all three of these quantities for randomly generated states. However, as seen in Fig.~\ref{fig:negvsbinegqubit}, the binegativity never differs too greatly from the negativity, which makes direct comparison of the three measures difficult to see in a three-dimensional plot. We can instead compare the concurrence to the differences $C-N$ and $N-N_2$, as shown in Fig.~\ref{fig:cvsnvsb}. In general, based on numerical evidence from randomly generated states, we conjecture that the the binegativity of two-qubit states is bounded by
\begin{equation}\label{eq:bnegbounds}
 \nu\frac{(c+\nu)(\nu+1)}{(c+\nu)^2+2c(1-c)}\leq N_2(\sigma)\leq \frac{\nu}{2}\frac{(c+\nu)^2}{c^2+\nu^2}
\end{equation}
for all states $\sigma$ with fixed negativity $N(\sigma)=\nu$ and concurrence $C(\sigma)=c$.  Furthermore, these bounds are actually achievable by states of the form given in \eqref{eq:sigpqr} below, as shown in Appendix \ref{app:bounds}. The lower and upper surfaces of the region shown in Fig.~\ref{fig:cvsnvsb} are derived from the lower and upper bounds in \eqref{eq:bnegbounds}. The binegativity of all randomly generated two-qubit states was found to be contained within the region. 

To derive the bounds conjectured in \eqref{eq:bnegbounds}, we define the following family of states:
\begin{equation}\label{eq:sigpqr}
 \sigma(p,q,r) := p\ketbra{\phi_{q}}{\phi_{q}} + (1-p)\ketbra{\psi_{r}}{\psi_{r}},
\end{equation}
 for $p,q,r\in[0,1]$, which are mixtures of the pure states
\begin{align*}
            \ket{\phi_q}&=\sqrt{q}\ket{00}+\sqrt{1-q}\ket{11}\\
\text{and } \ket{\psi_r}&=\sqrt{r}\ket{01}-\sqrt{1-r}\ket{10}.
\end{align*}
This family includes the states defined in \eqref{eq:sigMEMS}. Explicit computations of the binegativity, negativity, and concurrence of the states in \eqref{eq:sigpqr} are given in Appendix \ref{app:computations}. Here, however, we use this family to provide one example of a pair of states that have equal concurrence and equal negativity but different values of the binegativity. In particular, consider the following two states of the form $\sigma(p,q,r)$:
 \begin{equation}\label{eq:rho12}
 \begin{split}
  \rho_1&=\sigma\left(\frac{7}{48},1,\frac{1}{2}+\frac{\sqrt{1105}}{82}\right)\\
  \text{and }\,
  \rho_2&=\sigma\left(\frac{39}{112},\frac{1}{2}+\frac{2\sqrt{77}}{39},\frac{1}{2}\right).
 \end{split}
 \end{equation}
Straightforward computation shows that these states have negativity and concurrence given by $N(\rho_1)=N(\rho_2)=3/8$ and $C(\rho_1)=C(\rho_2)=1/2$. But the binegativities of these states differ, since 
 \begin{align*}
  N_2(\rho_1)=\frac{147}{400} =0.3675  
  \quad\text{and}
  \quad
  N_2(\rho_2)=\frac{77}{216} \approx 0.3564.
 \end{align*}
Furthermore, the binegativites of these states are exactly the lower and upper bounds conjectured in \eqref{eq:bnegbounds} for $c=1/2$ and $\nu=3/8$. These states are depicted as the ends of the vertical blue line in the centre of Fig.~\ref{fig:cvsnvsb}.


The above analysis shows the existence of states that have the same negativity and same concurrence, but have different values for the binegativity. If it turns out that the binegativity is indeed an entanglement measure, these states would be inequivalent with respect to LOCC even though they have the same values of negativity and concurrence. Since their negativities and concurrences coincide, it is impossible to  discern from the negativity and concurrence alone whether or not these states cannot be converted into each other via LOCC. The binegativity would yield an ordering of the entangled two-qubit states that is distinct from the ordering determined by the concurrence and the negativity. However, we currently only have numerical evidence that the binegativity is monotonic under PPT operations. It is therefore important to further investigate the binegativity to determine with certainty whether it is a true entanglement measure. \\

\emph{\textbf{Conclusion}} --- We investigated some properties of the binegativity $N_2(\sigma)=\Tr[\sigma^\Gammaminus]+2\Tr[\sigma^{\Gammaminus\Gammaminus}]$, which we conjecture to be entanglement measure for two-qubit states. The ordering that this new measure induces on the set of entangled two-qubit states was compared to that of the concurrence and the negativity. 

If $N_2$ turns out to be an entanglement measure, it would yield new insights toward understanding the structure of two-qubit entanglement, even though there is no known physical interpretation of the binegativity. While we have numerical evidence that $N_2$ is an entanglement measure of two-qubit states, we have no insight as to whether $N_2$ is an entanglement measure for states of larger systems, or whether it is a full entanglement monotone (i.e., if it is non-increasing on average for LOCC) \cite{Vidal2000,Plenio2005}. It would be worthwhile to investigate the binegativity and similar quantities for states of larger systems. Such analysis might yield further useful entanglement measures that can be used to investigate the structure of bipartite mixed entanglement in larger systems. \\


\begin{acknowledgments}
MWG is partially supported by Alberta Innovates--Technology Futures (AITF) and the Killam Trusts. GG is supported by NSERC.
\end{acknowledgments}


\bibliographystyle{mystyle2}
\bibliography{/home/mark/Documents/library.bib,extra.bib}


\appendix

\onecolumngrid

 

\section{Calculations}

\label{app:computations}

In this section, we compute the concurrence, negativity, and binegativity of the family of two-qubit states defined in \eqref{eq:sigpqr}. These states are defined by 
\begin{equation}\label{eq:appsigpqr}
 \sigma(p,q,r) := p\ketbra{\phi_{q}}{\phi_{q}} + (1-p)\ketbra{\psi_{r}}{\psi_{r}},
\end{equation}
 for $p,q,r\in[0,1]$, where
\begin{align*}
            \ket{\phi_q}&=\sqrt{q}\ket{00}+\sqrt{1-q}\ket{11}\\
\text{and } \ket{\psi_r}&=\sqrt{r}\ket{01}-\sqrt{1-r}\ket{10}.
\end{align*}
We can write these states in the standard basis as
\[
 \sigma= \begin{pmatrix}
                       pq & 0 & 0 & p\sqrt{q(1-q)}\\
                       0 &  (1 - p) r  & -(1-p)\sqrt{r(1-r)}  & 0\\
                       0 &  -(1-p)\sqrt{r(1-r)}  & (1 - p) (1-r)  & 0\\
                       p\sqrt{q(1-q)} & 0 & 0 &p(1-q)
                      \end{pmatrix}.
\]

We will show that the concurrence of the states $\sigma=\sigma(p,q,r)$ is given by
\begin{equation} \label{eq:appconc}
 C(\sigma)=2\sizedabs{p \sqrt{q (1-q)}-(1-p) \sqrt{r (1-r)}}.
\end{equation}
To compute the negativity and binegativity, we define 
\begin{equation}
 \alpha := p^2q(1-q) \quad\text{and}\quad \beta:=(1-p)^2r(1-r).
\end{equation}
The concurrence simplifies to $C(\sigma)=2\abs{\sqrt{\alpha}-\sqrt{\beta}}$, and the negativity of these states is given by
\begin{equation}\label{eq:appneg}
 N(\sigma)=\left\{\begin{array}{ll}
                          \sqrt{4(\beta-\alpha)+p^2}-p & \text{ if}\quad\alpha\leq\beta\\ \\
                          \sqrt{4(\alpha-\beta)+(1-p)^2}-(1-p)& \text{ if}\quad\alpha\geq\beta.
                         \end{array}
\right.
\end{equation}
The binegativity is $N_2(\sigma)=\frac{1}{2}N(\sigma) + 2\Tr[\sigma^{\Gammaminus\Gammaminus}]$, where $\Tr[\sigma^{\Gammaminus\Gammaminus}] = \frac{1}{2}\sqrt{\mu(1-\mu)}N(\sigma)$
and
\begin{equation}
 \mu=\left\{\begin{array}{ll}
  \frac{4\beta}{4\beta + \left(\sqrt{4(\beta-\alpha)+p^2}+p(1-2q)\right)^2} & \text{if}\quad\alpha\leq\beta\\ 
  \\
  \frac{4\alpha}{4\alpha + \left(\sqrt{4(\alpha-\beta)+(1-p)^2}+(1-p)(1-2r)\right)^2} & \text{if}\quad\alpha\geq\beta.
 \end{array}\right.
\end{equation}
With this notation, the binegativity of these states simplifies to
\begin{equation} \label{eq:appbineg}
 N_2(\sigma) = N(\sigma)\left(\frac{1}{2}+\sqrt{\mu(1-\mu)}\right).
\end{equation}

\subsection{Concurrence}

Recall that the concurrence of a mixed state of two qubits can be computed as $C(\sigma)=\max\{0,\lambda_1-\lambda_2-\lambda_3-\lambda_4\}$, where $\lambda_1,\lambda_2,\lambda_3,\lambda_4$ are the eigenvalues of  $\sqrt{\sqrt{\sigma}\sigma_y\otimes\sigma_y \overline{\sigma}\sigma_y\otimes\sigma_y \sqrt{\sigma}}\}$ in decreasing order and $\overline{\sigma}$ is the matrix whose entries are the complex conjugate of those in $\sigma$. Note that
\[
 \sigma_y\otimes\sigma_y \ket{00} = -\ket{11}, \quad  
 \sigma_y\otimes\sigma_y \ket{11} = -\ket{00}, \quad 
 \sigma_y\otimes\sigma_y \ket{01} = \ket{10}, \quad \text{and}\quad
 \sigma_y\otimes\sigma_y \ket{10} = \ket{01}.
\]
Hence $\sigma_y\otimes\sigma_y\ket{\phi_q} = \ket{\phi_{1-q}}$ and $\sigma_y\otimes\sigma_y\ket{\psi_r} = \ket{\psi_{1-r}}$, and thus 
\[
 \sigma_y\otimes \sigma_y\Bigl(p\ketbra{\phi_q}{\phi_q}+(1-p)\ketbra{\psi_r}{\psi_r}\Bigr)\sigma_y\otimes\sigma_y = 
 p\ketbra{\phi_{1-q}}{\phi_{1-q}}+(1-p)\ketbra{\psi_{1-r}}{\psi_{1-r}}.
\]
Now $\sqrt{\sigma}=\sqrt{p\ketbra{\phi_q}{\phi_q}+(1-p)\ketbra{\psi_r}{\psi_r}} = \sqrt{p}\ketbra{\phi_q}{\phi_q}+\sqrt{(1-p)}\ketbra{\psi_r}{\psi_r}$, so we need to compute the eigenvalues of 
\begin{align*}
 &\sqrt{\sqrt{\sigma}\sigma_y\otimes\sigma_y \overline{\sigma}\sigma_y\otimes\sigma_y \sqrt{\sigma}}=\\
 &= \sqrt{
 \Bigl(\sqrt{p}\ketbra{\phi_q}{\phi_q}+\sqrt{(1-p)}\ketbra{\psi_r}{\psi_r}\Bigr)
  \Bigl(p\ketbra{\phi_{1-q}}{\phi_{1-q}}+(1-p)\ketbra{\psi_{1-r}}{\psi_{1-r}}\Bigr)
 \Bigl(\sqrt{p}\ketbra{\phi_q}{\phi_q}+\sqrt{(1-p)}\ketbra{\psi_r}{\psi_r}\Bigr)
 } \\
 &= \sqrt{p^2\abs{\braket{\phi_q}{\phi_{1-q}}}^2 \ketbra{\phi_q}{\phi_q} + (1-p)^2\abs{\braket{\psi_r}{\psi_{1-r}}}^2\ketbra{\psi_r}{\psi_r}} \\
 &= p\abs{\braket{\phi_q}{\phi_{1-q}}} \ketbra{\phi_q}{\phi_q} + (1-p)\abs{\braket{\psi_r}{\psi_{1-r}}}\ketbra{\psi_r}{\psi_r}
\end{align*}
which are $2p\sqrt{q(1-q)}$ and $2(1-p)\sqrt{r(1-r)}$, since 
\[
 \braket{\phi_q}{\phi_{1-q}} = 2\sqrt{q(1-q)} 
 \quad\text{and}\quad
 \braket{\psi_r}{\psi_{1-r}} = 2\sqrt{r(1-r)}.
\]
This yields that the concurrence is 
\begin{equation}
 C(\sigma(p,q,r))=2\sizedabs{p\sqrt{q(1-q)}-2(1-p)\sqrt{r(1-r)}},
\end{equation}
as desired.

\subsection{Negativity}

The partial transpose of these states is given by
\[
 \sigma^\Gamma = \begin{pmatrix}
                       pq & 0 & 0 & -(1-p)\sqrt{r(1-r)}\\
                       0 &  (1 - p) r  &  p\sqrt{q(1-q)} & 0\\
                       0 &  p\sqrt{q(1-q)}  & (1 - p) (1-r)  & 0\\
                       -(1-p)\sqrt{r(1-r)} & 0 & 0 &p(1-q)
                      \end{pmatrix}
\]
which has eigenvalues
\begin{align*}
 \frac{p}{2}&\pm \frac{1}{2}\sqrt{p^2 + 4\bigl((1-p)^2r(1-r)-p^2q(1-q)\bigr)}\\
 \text{and} \quad \frac{1-p}{2}&\pm \frac{1}{2}\sqrt{(1-p)^2 +4\bigl(p^2q(1-q)- (1-p)^2r(1-r)\bigr)}.
\end{align*}
Hence the negativity of these states is given by
\begin{equation*}
 N(\sigma)=\left\{\begin{array}{ll}
                          \sqrt{4\bigl((1-p)^2r(1-r)-p^2q(1-q)\bigr)+p^2}-p & \quad\text{ if}\quad p^2q(1-q)\leq(1-p)^2r(1-r)\\ \\
                          \sqrt{4\bigl(p^2q(1-q)- (1-p)^2r(1-r)\bigr)+(1-p)^2}-(1-p)& \quad\text{ if}\quad p^2q(1-q)\geq(1-p)^2r(1-r),
                         \end{array}
\right.
\end{equation*}
as desired.

\section{States with same concurrence and negativity but different binegativity}
\label{app:bounds}

We can use the states defined in \eqref{eq:appsigpqr} to find a family of states that have fixed concurrence equal to $c$ and fixed negativity equal to $\nu$, but varying binegativity, given any fixed values $c$ and $\nu$. For any $p$ in the range
\begin{equation}
  p_{\min} \leq p \leq  p_{\max}
 \qquad\text{with}\qquad
 p_{\min} = \frac{c^2-\nu^2}{2\nu} 
 \quad\text{ and }\quad
 p_{\max}=\frac{c(\nu+1)}{c+\nu}-\frac{c+\nu}{2}
\end{equation}
consider $\sigma=\sigma(p,q_p,r_p)$ with $q_p$ and $r_p$ defined by
\begin{align*}
q_p &=\frac{1}{2} + \frac{\sqrt{c ^2- \nu ^2}}{2c}\frac{1}{p} \sqrt{\left(p-\frac{c-\nu}{2}\right) \left(p+\frac{c+\nu}{2}\right)}\\
\text{and}\quad
r_p &=\frac{1}{2} + \frac{\sqrt{c ^2- \nu ^2}}{2c}\frac{1}{1-p} \sqrt{\left(p-\frac{c-\nu}{2}-\frac{c(\nu+1)}{c-\nu}\right) \left(p+\frac{c+\nu}{2}-\frac{c(\nu+1)}{c+\nu}\right)}.
\end{align*}
We can compute the concurrence, negativity, and binegativity of these states by plugging these values for $p$, $q$, ad $r$ into the formulas in \eqref{eq:appconc}, \eqref{eq:appneg}, and \eqref{eq:appbineg}. We find that the concurrence and negativity of these states are given by $C(\sigma) = c$, $N(\sigma) = \nu$. The binegativity of these states is
\begin{equation}\label{eq:neg2pqr}
 N_2(\sigma)  = \frac{\nu(c+\nu)}{4c}\left(2+\frac{c-\nu}{p+\nu}\right),
\end{equation}
which varies with $p$ as long as $c\neq \nu$. This allows us to find states that have the same concurrence and same negativity but different binegativity. The example in the main body of the paper is produced by picking 
\[
 \rho_1 = \sigma(p,q_p,r_p) \quad\text{with} \quad p=p_{\max}
 \qquad\text{and}\qquad
 \rho_2 = \sigma(p,q_p,r_p) \quad\text{with} \quad p=p_{\min}
\]
where $c=\frac{1}{2}$ and $\nu=\frac{3}{8}$.

For all the states of the form in \eqref{eq:appsigpqr}, the binegativity is bounded by
\begin{equation}\label{eq:appconc1}
 \frac{\nu}{2}\frac{(c+\nu)(\nu+1)}{(c+\nu)^2+2c(1-c)}\leq N_2(\sigma)\leq \frac{\nu}{2}\frac{(c+\nu)^2}{c^2+\nu^2},
\end{equation}
whenever $N(\sigma)=\nu$ and $C(\sigma)=c$. In fact, picking $p=p_{\min}$ and $p=p_{\max}$ yield the upper and lower bounds \eqref{eq:appconc1}. Since the upper and lower bounds are distinct whenever $\nu_c<\nu<c$, this implies the existence of states that have the same negativity and same concurrence, but have different values for the binegativity. This observation would have interesting consequences if it were to be proven that the binegativity is in fact an entanglement measure. In particular, these states would be inequivalent with respect to LOCC, but they have the same values of negativity and concurrence, so it would be impossible to  discern from the negativity and concurrence alone that these states cannot be converted into each other via LOCC. The binegativity therefore yields an ordering the entangled states of two-qubits that is distinct from the ordering determined by the concurrence and the negativity. It is therefore of great interest to further investigate the binegativity to see if it is indeed an entanglement measure.

 \end{document}